\documentclass[twocolumn,grl]{agutex}

 \usepackage{lineno}

\usepackage{xcolor}

\usepackage{amsmath}
\usepackage{indentfirst}
\ifx\pdfoutput\undefined
\usepackage[dvips]{graphicx}
\else
\usepackage[pdftex]{graphicx}
\usepackage{epstopdf}
\fi

\setkeys{Gin}{draft=false}
\authorrunninghead{AN ET AL.}

\titlerunninghead{WHISTLER WAVE EXCITATION}

\authoraddr{Corresponding author: X. An,
Department of Atmospheric and Oceanic Sciences, University of
California, Los Angeles.
(xinan@atmos.ucla.edu)}

\begin{document}

%
%

\title{Resonant excitation of whistler waves by a helical electron beam}
%
%

%
%



\authors{X. An\altaffilmark{1},
 B. Van Compernolle\altaffilmark{2}, J. Bortnik\altaffilmark{1},
 R. M. Thorne\altaffilmark{1}, L. Chen\altaffilmark{3}, W. Li\altaffilmark{1}}

\altaffiltext{1}{Department of Atmospheric and Oceanic Sciences,
University of California, Los Angeles, CA, USA.}

\altaffiltext{2}{Department of Physics, University of California,
Los Angeles, CA, USA.}

\altaffiltext{3}{Physics Department, W. B. Hanson center for 
Space Sciences, University of Texas at Dallas, TX, USA.}

%
%

\keypoints{
\item Chorus-like whistler mode waves are excited in a laboratory plasma to study excitation process.
\item The resonance structure of whistler wave is experimentally resolved for the first time.
\item Linear theory shows consistent behavior in both intensity and wave normal angle.
}



%
%


\begin{abstract}
Chorus-like whistler-mode waves that are known to play a fundamental role in driving radiation-belt dynamics are excited on the Large Plasma Device by the injection of a helical electron beam into a cold plasma. The mode structure of the excited whistler wave is identified using a phase-correlation technique showing that the waves are excited through a combination of Landau resonance, cyclotron resonance and anomalous cyclotron resonance. The dominant wave mode excited through cyclotron resonance is quasi-parallel propagating, whereas wave modes excited through Landau resonance and anomalous cyclotron resonance propagate at oblique angles that are close to the resonance cone. An analysis of the linear wave growth rates captures the major observations in the experiment. The results have important implications for the generation process of whistler waves in the Earth's inner magnetosphere.
\end{abstract}

%
%

%

\begin{article}

%
%

\section{Introduction}
Naturally-generated whistler mode chorus waves in the Earth's near-space environment are known to play a key role in the acceleration and loss of the relativistic electrons that comprise the radiation belts \citep[e.g.,][]{Thorne2010, thorne2013rapid}. They appear in various forms, such as discrete rising or falling tones, and broadband hiss emissions \citep{pope1963chorus, Burtis1968chorus, corn1978review, koons1981chorus, santolik2009chorus, liwen2012chorus}. These chorus waves are commonly separated into two distinct bands with a gap at $0.5\Omega_e$ ($\Omega_e$ is the electron gyrofrequency) \citep{Tsurutani1974chorus, Hayakawa1984chorus, Liwen2013chorus}. The lower band ($\omega/\Omega_e < 0.5$, $\omega$ is the wave frequency) is dominantly parallel-propagating with some portion close to the resonance cone \citep{agapitov2013chorus,Liwen2013chorus}, whereas the upper band ($\omega/\Omega_e > 0.5$) has a broad range of wave normal angles between $0^o$ and $60^o$ \citep{Liwen2013chorus}. Several mechanisms have been proposed to account for the banded structure of chorus \citep{Omura2009bandchorus, Liu2011bandchorus, Fuxr2014bandchorus, Mourenas2015bandchorus, Fuxr2015bandchorus}.
Among the proposed mechanisms, \citet{Mourenas2015bandchorus} suggested that the less frequently occurring but statistically significant very oblique lower band chorus waves can be generated through a combination of cyclotron resonance and Landau resonance with low energy electron beams having an energy of a few keV. But these theoretical ideas remain to be tested by satellite observations and laboratory experiments.

In order to study the excitation of chorus-like whistler waves in a laboratory plasma, an electron beam is used as the free energy source, injected into a cold background plasma. In a beam-plasma system, various instability processes can excite a variety of waves including whistler-mode waves \citep{Bell1964whist}, Bernstein-mode waves \citep{Kusse1970bernstein,Mizuno1971441} and Langmuir waves \citep{oneil1971smallbeam,Gentle1973beamexp}. Whistler-mode emissions by beam-plasma interaction have been studied extensively in the past, such as in the generation of auroral hiss \citep{Maggs1976hiss,Gurnett1983hiss,Sazhin1993153}, in active experiments in the space environment \citep{Lavergnat1979multibands,tokar1984beam,Gurnett1986spacelab2,Farrell1988spacelab2,Neubert1992153} and in controlled laboratory settings \citep{Stenzel1977beamhiss,Krafft1994Landau, Staro1999cyclotron, Staro1999transition,PRL_Chorus}. During active experiments of the Spacelab $2$ mission, for instance, beam-generated whistler-mode emissions were observed to propagate near the resonance cone and were attributed to Landau resonance \citep{Gurnett1986spacelab2,Farrell1988spacelab2}.
In controlled laboratory experiments, whistler-mode emissions were also generated through both Landau resonance \citep{Krafft1994Landau} and cyclotron resonance \citep{Staro1999cyclotron} using a density-modulated electron beam. However, if the electron beam is not modulated, broadband whistler waves are produced instead of a single wave with predetermined frequency. The experiment reported in this Letter is unique in that there is no imposed frequency on the electron beam. Whistler waves are spontaneously excited by different resonance modes simultaneously, which results in frequency spectra having clear upper and lower bands.

\section{Experimental setup}
\begin{figure}
 \centering
 \noindent\includegraphics[width=6.75in]{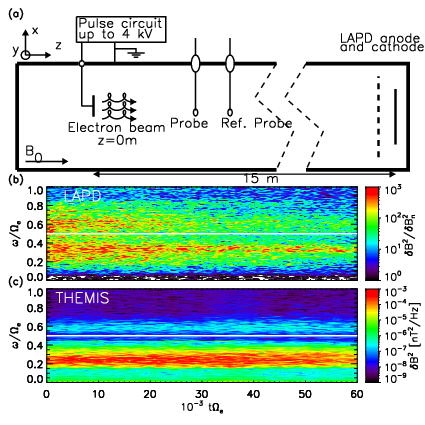}
 \caption{(a) A schematic diagram of the experimental setup. A $10$ cm diameter electron source launches an electron beam with energy up to $4$ keV. Probes measure the plasma parameters and detect wave activity. A reference probe is added to construct the wave mode structure. (b) A typical wave spectrogram taken from LAPD experiment, showing the two-band structure with a gap at $0.5 \Omega_e$. Note that $\delta B_n$ is the spectral density of magnetic noise level as a function of frequency. $(\delta B / \delta B_n)^2$ measures the amplification of magnetic field from noise level. (c) A typical wave spectrogram taken from the THEMIS-A satellite at $\mbox{L}=7.7-7.8$ and $\mbox{MLT}=7.2$ near the magnetic equator, with a clear upper and lower band separated by $0.5 \Omega_e$.}
 \label{fig1}
 \end{figure}

The experiment is performed on the upgraded Large Plasma Device (LAPD) \citep{lapd,lapdsource} at the Basic Plasma Science Facility (BaPSF) at University of California, Los Angeles. The LAPD is a long cylindrical device, with an axial magnetic field and an 18 m long, 60 cm diameter quiescent plasma column. A schematic diagram of the experimental setup is shown in Figure \ref{fig1}a. The background plasma is pulsed at 1 Hz. A 10 cm diameter electron beam source ($0.5 \ \mbox{kV} \leq V_{beam} \leq 4 \ \mbox{kV}$) \citep{PRL_Resonance_condition,PRL_Chorus}  is introduced into the machine 15 m from the LAPD source at the opposite end of the machine. The beam source is angled at $30^o$ with respect to the background magnetic field in order to provide sufficient free energy in the electron distribution for the cyclotron growth of the whistler waves. The magnetic field near the beam source is uniform at $60$ G for $7$ meters, and then transitions to $350$ G near the LAPD source for reliable operation of the source. The Helium plasma in the experiment has a Helium fill pressure of $3 \times 10^{-5}$ Torr. The experiment is performed after the active phase of the LAPD discharge, in the afterglow plasma, when the electrons are relatively cold, i.e., $T_e \leq 0.5$ eV.  Typical absolute plasma parameters in the laboratory are quite different from those found in the magnetosphere but the dominant scaled dimensionless quantities are set to be very similar (see table I of \citet{PRL_Chorus} for the range of plasma parameters). Measurements of plasma parameters and wave activity were taken with Langmuir probes and high frequency magnetic loop probes (which is positioned $1$ meter away from the beam source). Volumetric data is obtained by moving computer controlled probes through the plasma over the course of thousands of identical plasma shots.  The start of the electron beam pulse is taken as $t = 0$ and the streaming direction of the electron beam as positive direction of $z$.
Figure \ref{fig1}b shows an example of a dynamic spectrogram of the whistler waves for a $3 \ \mbox{keV}$ beam firing into a cold plasma with $\omega_{pe}/\Omega_e = 9.6$, $n_b/n_0 = 1.6 \times 10^{-3}$, $n_0 = 3.2 \times 10^{10} \mbox{cm}^{-3}$, where $\omega_{pe}$ is the electron plasma frequency, $n_b$ is the beam density and $n_0$ is the total plasma density. Whistler waves are present in a broad frequency range with $0.1<\omega/\Omega_e<0.9$. This example is designated as the nominal case, which is investigated in detail in Section \ref{kz-omega}.  For comparison, a spectrogram of whistler waves at $\mbox{L}=7.7-7.8$ from the THEMIS-A satellite \citep{angelopoulos2009themis} is shown in Figure \ref{fig1}c, with $\omega_{pe}/\Omega_e=5.2$. The similarity between the observation made in space, and that made in the laboratory is immediately apparent. The incoherent two-band structure in the experiment closely resembles the two-band structure of whistler mode chorus waves in space, which will be demonstrated below to be produced by a number of different resonance modes.

\section{$k_z-\omega$ diagram}
\label{kz-omega}
\begin{figure}
   \centering
   \includegraphics[width=6.75in]{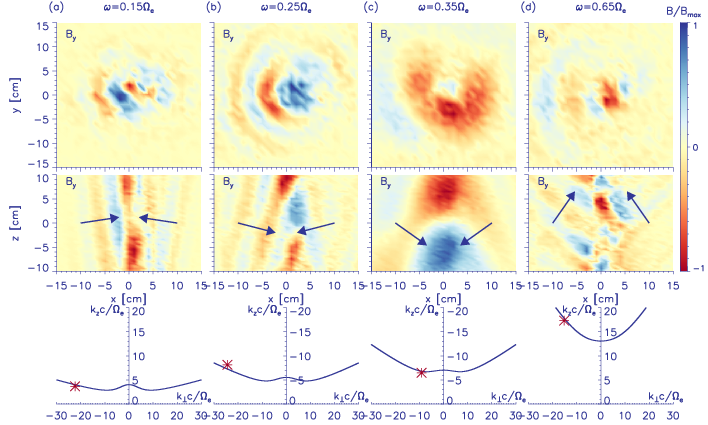}
   \caption{(color online). Mode structure of whistler waves at $4$ representative frequencies corresponding to each column. The first two rows show $B_y$ in the $x-y$ and the $x-z$ planes, respectively. Wave amplitudes are normalized to the maximum wave amplitude in each panel. Arrows in the second row represent the wave vector direction. The third row shows the refractive index surface (which is a curve in $2$D projection) for each frequency in the wave number space. The wave number corresponding to each mode structure is marked by the red asterisk. Note that $k_\perp<0$ represents radially inward propagating waves. An animation of each of these cases is available online as part of the supplemental material.}
   \label{fig2}
 \end{figure}

Using a moving magnetic field probe together with a similar fixed probe as a reference, the phase-delay $\Delta\phi(\mathbf{x},\omega)$ between these two probes was measured accurately using a cross-correlation technique. Together with the measured amplitude $\big| \tilde{B}(\mathbf{x},\omega) \big|$, the $3$D mode structure can be constructed as $\big| \tilde{B}(\mathbf{x},\omega) \big| \exp \big( i\Delta\phi(\mathbf{x},\omega) -i\omega t \big)$ using data from many identical plasma shots. Figure \ref{fig2} shows the mode structure and corresponding refractive index surfaces of broadband whistlers at $4$ representative frequencies (animations of the wave propagation at various frequencies are available in the supplemental material). From the mode structure, one can extract both parallel and perpendicular wave numbers. The phase-delay between the moving probe and the reference probe in the cylindrical geometry is assumed to be $\Delta\phi(\mathbf{x},\omega) = k_z(\omega) z + k_\perp(\omega) \rho$, which has been obtained in the process of constructing the mode structure. $k_z$ and $k_\perp$ are the parallel and perpendicular wave numbers, respectively. $z$ and $\rho$ are the axial and radial coordinates in the cylindrical geometry, respectively. To extract the parallel wave number, for a given frequency, one can linearly fit $\Delta\phi(\rho,z,\omega)$ in the $x-z$ plane as a function of $z$ at each $\rho$ position. The perpendicular wave number is obtained, in a similar fashion, by fitting $\Delta\phi(\rho,z,\omega)$ as a linear function of $\rho$ at each $z$ position. The parallel wave number and perpendicular wave number for each mode structure is marked using the red asterisk in the wave number space in the $3$rd row in Figure \ref{fig2}.
The dispersion relation of whistler waves in a cold plasma is,
\begin{linenomath}
\begin{eqnarray}
n^2 = \dfrac{k^2 c^2}{\omega^2} = \dfrac{\omega_{pe}^2}{\omega(\Omega_e \cos \psi - \omega)}
\end{eqnarray}
\end{linenomath}
Here $n$ is the refractive index. Wave normal angle (WNA) $\psi$ is used to denote the angle between the wave vector and the background magnetic field. The term resonance cone is used to describe a region in the wave number space, where whistler wave propagation is prohibited ($n^2<0$), approximately where $\psi > \arccos(\omega/\Omega_e)$. More generally, whistler waves are strongly damped in the resonance cone region when thermal effects of the background plasma are taken into account \citep{HORNE1990311}.
The mode structure at $\omega=0.15\Omega_e$ is displayed in column (a). The wave fronts propagate towards the center, where the beam is located, in the $x-y$ plane (top panel). These features can be understood by locating the $\mathbf{k}$-vector in wave number space. The group velocity $\partial\omega/\partial\mathbf{k}$ is normal to the refractive index surface. In this case the perpendicular group velocity is directed in an opposite sense to the perpendicular phase velocity, which means energy is flowing out and away from the beam. Consistently, these waves propagate close to the resonance cone with WNA$\approx 80^o$ in the $x-z$ plane. The co-streaming of these waves with the electron beam in the parallel direction indicates that the excitation is likely due to Landau resonance. The mode structure at $\omega=0.25\Omega_e$ is displayed in column (b). In contrast to the waves at $\omega=0.15\Omega_e$, these waves are counter-streaming with the beam, having WNA$\approx 105^o$, which is also close to resonance cone. It will be shown below that the excited waves at this frequency are due to first-order cyclotron resonance. The mode structure at $\omega=0.35\Omega_e$ is displayed in column (c). This mode has slightly oblique wave fronts whereas the group velocity is parallel to the background magnetic field, known as Gendrin mode. This is the location of the spectral peak shown earlier in panel (b) of Figure \ref{fig1}, and hence constitutes the most intense waves in this experiment. Waves travelling in the opposite direction to the electron beam can efficiently gain energy from the beam through cyclotron resonance, resulting in wave amplitudes that are larger than that of other modes. The mode structure at $\omega=0.65\Omega_e$ is displayed in column (d). These waves are co-streaming with the electron beam with WNA$\approx 49^o$. Animation of $B_x-B_y$ vector fields show the co-existence of both a strong right-hand polarized component and a weak left-hand polarized component. This mode is more complicated than the previous ones in the sense that it is excited through both Landau resonance and anomalous cyclotron resonance, as shown below.

\begin{figure}
 \centering
 \noindent\includegraphics[width=6.75in]{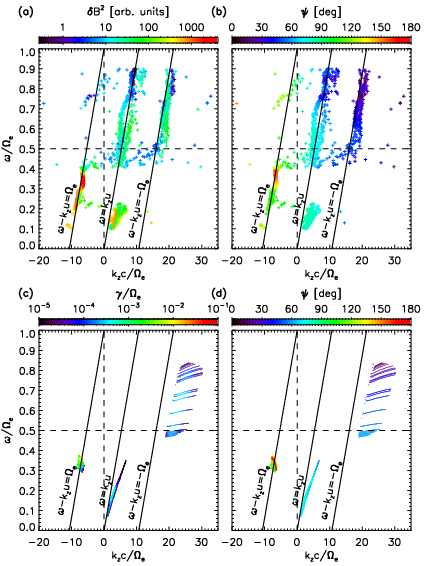}
 \caption{Wave properties plotted on a $k_z-\omega$ diagram from the LAPD experiment (a, b) and corresponding HOTRAY calculations (c, d), respectively, showing multiple resonance modes, color-coded by (a) power spectral density, (b) wave normal angle $\psi$ from the experiment, (c) linear growth rates and (d) wave normal angle $\psi$ from HOTRAY code.}
 \label{fig3}
 \end{figure}

Using the method to extract the wave numbers described above and shown in Figure \ref{fig2}, we repeat the procedure for all frequencies in the range $0.1<\omega/\Omega_e<0.9$, and show the corresponding results in panels (a) and (b) of Figure \ref{fig3}. The data is organized in the $k_z-\omega$ plane in panel (a), color coded by the power spectral density. Panel (b) is organized in the same way as panel (a), but color coded by WNA computed as $\psi = \arctan(k_\perp/k_z)$. 
These data points beautifully aggregate into three resonance modes, corresponding to $\omega - k_z u = n\Omega_e$ with $n=1,0,-1$ ($u$ is the initial beam velocity in the parallel direction). Cyclotron resonance occurs in the frequency range below $0.4 \Omega_e$ as shown by the cluster of orange-red points falling on the $n=1$ line. Even for normal, first-order cyclotron resonance, the WNA is seen to be close to resonance cone in the frequency range $0.2<\omega/\Omega_e<0.3$, which becomes less oblique with increasing frequency and eventually reaches about $160^o$ in the most intense range between $0.35<\omega/\Omega_e<0.4$. The cyclotron resonance mode results in the most efficient energy transfer and hence the largest wave amplitude compared to the other two resonance modes. For Landau resonance below $0.25 \Omega_e$, the measured data is seen to fall slightly below $\omega=k_z u$, which may be evidence that the beam electrons are slowed down from their initial velocity $u$, by the time they reach the primary wave excitation region in the experiment. The Landau resonance mode below $0.25 \Omega_e$ has a WNA of $80^o$ or so. Such oblique WNA implies a finite wave electric field (though small) in the parallel direction, which allows energy transfer between beam electrons and whistler waves through Landau resonance. The WNA for Landau resonance in the frequency range $0.4<\omega/\Omega_e<0.9$ is near the edge of the resonance cone. Anomalous cyclotron resonance co-exists with Landau resonance in the frequency range $0.4<\omega/\Omega_e<0.9$. The WNA for anomalous cyclotron resonance is in the range from $20^o$ to $50^o$.

To compare the experimental results with the predictions of linear theory, the hot plasma dispersion relation is solved to obtain linear growth rates and the accompanying WNA using the HOTRAY code \citep{JGRA:JGRA9431}. This code implements the hot plasma dispersion relation in a magnetized plasma for an arbitrary number of summed Maxwellian distributions including an optional drift in the parallel direction and also a loss cone. The beam electrons are modeled as a beam ring distribution, implemented in HOTRAY as
\begin{linenomath*}
\begin{eqnarray}
f(v_\perp , v_z) = \dfrac{1}{\pi^{\frac{3}{2}}\alpha_\perp^2\alpha_\parallel(1-\beta)} \exp\left( -\frac{(v_z-v_d)^2}{\alpha_\parallel^2} \right) \nonumber \\
\times \Bigg( \exp\left(-\frac{v_\perp^2}{\alpha_\perp^2}\right) - \exp\left(-\frac{v_\perp^2}{\beta\alpha_\perp^2}\right) \Bigg) 
\label{distribution}
\end{eqnarray}
\end{linenomath*}
Here $v_\perp$ and $v_z$ are perpendicular and parallel velocity, respectively, relative to the background magnetic field. $\alpha_\parallel$, $\alpha_\perp$, $\beta$ and $v_d$ are free parameters that control the shape of the distribution function. Since direct measurements of the distribution function are not available at this stage, the distribution function is roughly inferred based on physical arguments. As the beam enters the plasma, the fastest growing Langmuir waves slow down and relax the beam electrons in the parallel direction \citep{oneil1971smallbeam,Gentle1973beamexp}. The beam electrons move locally over a single Langmuir wave with a relative velocity of $\Delta v = 2^{-\frac{4}{3}} (n_b/n_0)^{\frac{1}{3}} u$ \citep{oneil1971smallbeam}. The Langmuir wave eventually reaches an amplitude  $\phi \approx m_e (\Delta v)^2 /e$ which is enough to trap the beam electrons, and also causes nonlinear saturation of wave growth. Thus the beam electrons are modeled to be a Maxwellian centered at the phase velocity of the Langmuir wave $v_d = u-\Delta v$ with a thermal spread $\alpha_\parallel = \sqrt{2}\Delta v$. In the perpendicular direction, we set $\alpha_\perp = v_{\perp0}$ and $\beta = 0.8$. Here $v_{\perp0}$ is the initial beam velocity in perpendicular direction. As such, the corresponding distribution in the perpendicular direction peaks at $(v_\perp)_{\rm max} = v_{\perp0}$ and the full width at half Maximum is $\sim v_{\perp0}$, which is likely much broader than that in the experiment, but is the lower limit that can be reached by equation \eqref{distribution}. The results of solving the hot plasma dispersion relation for the distribution described above are shown in panels (c) and (d) of Figure \ref{fig3}, in the same format as the first two panels of Figure \ref{fig3}. The cyclotron resonance mode, Landau resonance mode and anomalous cyclotron resonance mode show up in approximately the right $k_z-\omega$ locations, with appreciable growth rates, and having consistent WNA with the experimental results. The dominant wave mode is excited through cyclotron resonance when the resonant velocity is at the negative gradient side $\partial f/\partial v_z<0$ \citep{JGR:JGR5503, kennel1966whist}. Thus the calculated data points for cyclotron resonance have a slope larger than $v_d$ but near the initial beam velocity $u$.
The Landau resonance mode, on the other hand, is excited by the positive gradient $\partial f/\partial v_z>0$. Therefore the calculated data points for Landau resonance have a slope below $v_d$. A broad spectrum between $0.4 < \omega/\Omega_e < 0.85$ is excited through anomalous cyclotron resonance with WNA close to resonance cone. We note that waves in Landau resonance with beam electrons in the frequency range $\omega/\Omega_e > 0.4$ are not well captured by the linear growth rate calculations.
However, one should also note that in reality spatial wave growth takes place. Waves with $k_z<0$ ($k_z>0$) are growing toward (away from) the injection point of the beam electrons, whereas we make a simplifying assumption and construct one distribution function to account for all waves with different profiles of spatial growth. This very likely results in some inconsistency between the linear analysis and experimental results, although the general trends are remarkably similar.

\section{Parameter scans}
To investigate whether the linear growth rates scale in a similar way as the observed waves under a variety of plasma conditions, we perform a series of parameter scan. The plasma parameters used in Figure \ref{fig1}b were chosen to serve as the control case, and the beam source was tilted to $45^o$. For clarification, the parameters for the control case are $\omega_{pe}/\Omega_e = 9.6$, $E_b = 3$keV, $n_b/n_0 = 1.6 \times 10^{-3}$ and pitch angle $\alpha = 45^o$. For a given frequency, a calculation of linear growth rates is performed over all possible WNAs and the largest linear growth rate is extracted. Repeating this procedure for many frequencies in the whistler wave range gives a spectrum of linear growth rate. Thus a comparison between power spectral density and linear growth rate can be made for different plasma parameters.

\begin{figure}
 \centering
 \noindent\includegraphics[width=6.75in]{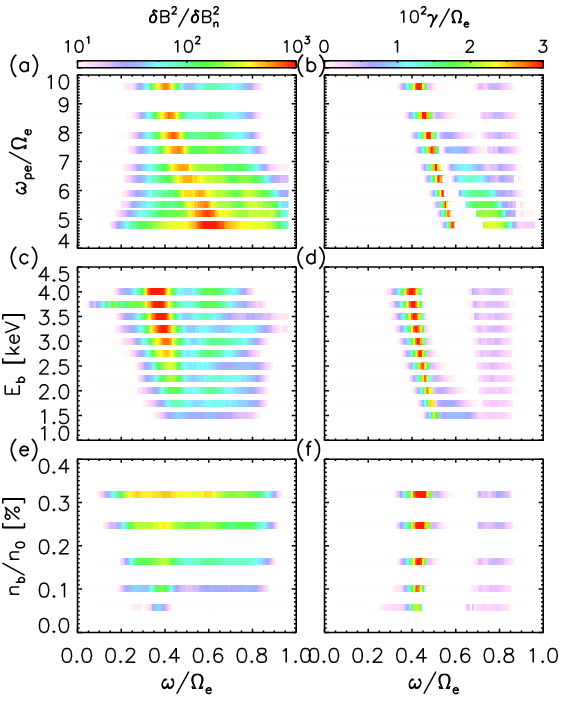}
 \caption{A comparison of observed wave properties (panels a, c, and e) and corresponding maximum linear growth rates (panels b, d, and f) obtained from three parameter scans: variation of the plasma density (a, b), beam energy (c, d) and beam density (e, f). $\delta B_n$ is the spectral density of magnetic noise level as a function of frequency. $(\delta B / \delta B_n)^2$ measures the amplification of magnetic field from noise level.}
 \label{fig4}
 \end{figure}

The first parameter scan was performed by varying the cold plasma density at a fixed beam energy $E_b=3$ keV and fixed beam density $n_b= 5 \times 10^7 cm^{-3}$. This changes both $\omega_{pe}/\Omega_e$ and $n_b/n_0$. Figure \ref{fig4}a shows power spectral densities for different $\omega_{pe}/\Omega_e$ from the experiment. The corresponding linear growth rates from HOTRAY are shown in Figure \ref{fig4}b. The power spectral peaks clearly shift to higher frequency as the plasma density decreases, which is well captured by the spectra of linear growth rate. These spectral peaks are parallel propagating wave modes and are dominantly excited through normal cyclotron resonance, described by $\omega-k_zu=\Omega_e$.
An up-shift in the wave frequency is required to satisfy the resonance condition as the plasma density decreases. The very oblique wave modes excited through Landau resonance also show an up-shift in frequency as shown in Figure \ref{fig4}b in yellow-green bands.
The wave modes excited through anomalous cyclotron resonance and higher order resonance are relatively insensitive to plasma density changes,  ranging between $0.7-0.9 \Omega_e$.
A second scan, displayed in Figure \ref{fig4}c and \ref{fig4}d, was done at $\omega_{pe}/\Omega_e=9.6$ and $n_b/n_0 = 1.6 \times 10^{-3}$ by varying electron beam energy. An up-shift of the power spectral peak is seen as beam energy decreases (Figure \ref{fig4}c), which agrees well with linear analysis (Figure \ref{fig4}d). It can be understood by observing that, to satisfy $\omega-k_zu=\Omega_e$, the wave frequency has to increase to compensate for the reduction of the Doppler shift term. The secondary peaks in the upper band $\omega/\Omega_e>0.5$ in Figure \ref{fig4}c, primarily due to Landau resonance, are not well reproduced by the linear analysis.
A third scan was performed by varying beam density at $\omega_{pe}/\Omega_e=9.6$ and $E_b=3$ keV as shown in Figure \ref{fig4}e and \ref{fig4}f. Both the power spectral density and linear growth rate increase as $n_b/n_0$ increases, which is expected. It is noted that there is frequency spectrum broadening in Figure \ref{fig4}e (also in Figure \ref{fig4}a) as $n_b/n_0$ increases. Preliminary investigation shows that wave-wave interactions get enhanced as $n_b/n_0$ increases. This possibly leads to the broadening of power spectrum. Another possibility is the broadening of wave-particle resonance resulting from strong turbulence \citep{dupree1966}.

\begin{figure}
 \centering
 \noindent\includegraphics[width=6.75in]{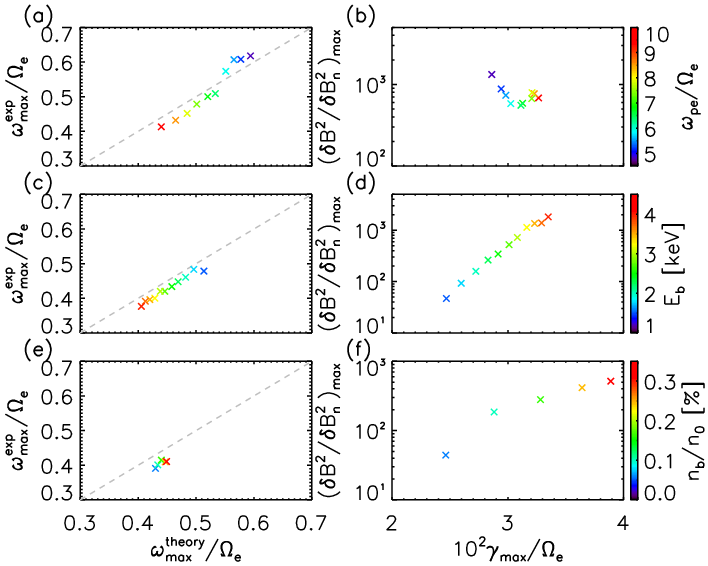}
 \caption{A comparison of the spectral peak frequencies from experiment and linear theory (a, c, e), and also comparisons of maximal saturated wave power with maximal linear growth rate (b, d, f) for three parameter scans. Comparisons for plasma density scan are displayed in (a, b), beam energy scan in (c, d), and beam density scan in (e, f), for which the corresponding parameters are plotted as color-bars at the right edge.}
 \label{fig5}
 \end{figure}

Finally, the peak of each wave power spectrum from the experiment and that of each linear growth rate spectrum are extracted for all parameter scans. This serves to compare the spectral peak locations from both experiment and linear theory (Figure \ref{fig5}a, \ref{fig5}c, \ref{fig5}e), and also to compare the maximal saturated wave power with the maximal linear growth rate (Figure \ref{fig5}b, \ref{fig5}d, \ref{fig5}f). Three rows in Figure \ref{fig5}, from top to bottom, correspond to plasma density scan, beam energy scan and beam density scan, respectively. Comparisons of the spectral peak frequencies show good agreement between experiment and linear theory. The shift of the spectral peak frequencies with respect to plasma density and beam energy can be understood in terms of linear excitation through cyclotron resonance as explained above. The shift of the spectral peak frequencies with respect to beam density is due to the modification of plasma dispersion relations by hot electrons.
One may expect some positive correlation between saturated wave power and linear growth rate, though the wave saturation process is governed by nonlinear processes. This expectation is consistent with the beam energy scan and the beam density scan. However, the saturated wave power and the linear growth rate show an inverse relation in the low plasma density regime. This indicates that nonlinear processes become a stronger factor in mediating wave-particle interactions as $n_b/n_0$ gets larger.

\section{Summary}
In summary, we have used a novel experimental setup to reveal the complex mode structure and excitation mechanisms of whistler-mode waves in a laboratory plasma, designed to closely resemble the plasma characteristics in the near-Earth space environment. Our results show that the whistler waves are excited primarily due to three basic resonance regimes simultaneously: the normal cyclotron resonance mode, Landau resonance, and first order anomalous cyclotron resonance mode. Linear wave growth calculations show consistent behavior both in intensity (or growth rate) and wave normal angle, shedding new light on the excitation process of whistler waves in space. However, a bi-Maxwellian distribution is observed to be responsible for the excitation of whistler-mode chorus waves in space, whereas our experiment essentially used a beam ring distribution in velocity space. The difference in the distribution function of hot electrons leads to some of the different wave characteristics between the laboratory experiment and space observations. The diagnostics on electron distribution function are also desired. These issues are the directions of future studies.


%
%
%
%
%
%
%

\begin{acknowledgments}
The authors wish to thank W. Gekelman, G. J. Morales and T. A. Carter for insightful discussions. The research was funded by the
Department of Energy and the National Science Foundation by Grant No. DE-SC0010578, which was awarded to UCLA through the NSF/DOE Plasma Partnership program. Work was done at the Basic Plasma Science Facility also funded by DOE/NSF.
\end{acknowledgments}

\end{article}
%
%
%
%
%
%
\newpage
 
%
%


\end{document}